\documentclass[5p,number,letterpaper]{elsarticle}
\newcommand{\Lor}{\mathcal{L}_0}

\usepackage{amssymb}
\usepackage{amsmath}

\newcommand{\q}{{\mathbf q}}

\journal{Solid State Communications}

\begin{document}

\begin{frontmatter}

\title{Electron-electron interaction induced spin thermalization in quasi-low-dimensional spin valves}

\author{Tero T. Heikkil\"a}

\address{Low Temperature Laboratory, Aalto University School of Science and Technology, FI-00076 AALTO, Finland}

\author{Moosa Hatami}
\author{Gerrit E. W. Bauer}
\address{Kavli Institute of NanoScience, Delft University of
Technology, 2628 CJ Delft, The Netherlands}

\begin{abstract}
We study the spin thermalization, i.e., the inter-spin energy
relaxation mediated by electron-electron scattering in small spin
valves. When one or two of the dimensions of the spin valve spacer
are smaller than the thermal coherence length, the direct spin
energy exchange rate diverges and needs to be regularized by the
sample dimensions. Here we consider two model systems: a long
quasi-1D wire and a thin quasi-2D sheet.
\end{abstract}


\end{frontmatter}

\section{Introduction}
\label{sec:intro}

The thermoelectric response of a ferromagnet$|$normal
metal$|$ferromagnet spin valve \cite{hatami07,hatami09,dubi09}
depends sensitively on the strength of inter-spin energy relaxation
(spin thermalization) inside the normal metal spacer
\cite{heikkilaup09}. In large structures at high temperatures, spin
thermalization is dominated by electron-phonon coupling, whereas at
low temperatures direct spin-flip scattering becomes important. A
third mechanism is the electron-electron scattering, which in
relatively large spin valves is weak and can typically be neglected.
However, for smaller spin valves the electron-electron interaction
becomes stronger. When one or more of the dimensions are smaller
than the thermal coherence length $\xi_T=\sqrt{\hbar D/(k_B T)}$,
the kernel of the electron-electron collision integral should be
calculated for reduced (1D or 2D) spatial dimensions. In these cases
the thermalization rate formally diverges
\cite{dimitrova07,chtchelkatchev08} and needs to be properly
regularized. In this paper we discuss such regularization schemes
and calculate the resulting thermalization rates.

\section{Theory}
\label{sec:theory} In a biased spin valve with an antiparallel
configuration of the magnetic reservoirs, the electron distribution
function may depend on the spin index $\sigma$. In this case the
electron-electron interactions can be described by the three
different collision integrals represented by the diagrams in
Fig.~\ref{fig:eecoll} and calculated as explained in the Appendix.

\begin{figure}[h]
\centering
\includegraphics[width=\columnwidth]{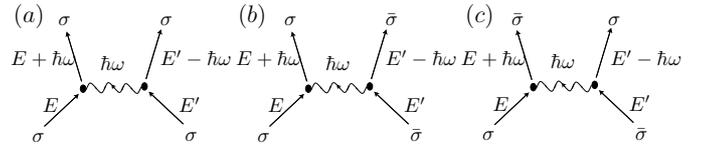}
\caption{The three types of electron-electron scattering vertices in
a system with spin-dependent distribution functions $f_\sigma(E)$.
a) Equal-spin scattering that does not lead to spin thermalization,
b) spin conserving scattering and c) spin exchange scattering. The
latter two cause spin thermalization and are discussed in this
paper.} \label{fig:eecoll}
\end{figure}

The kernels of the collision integrals depend on the Fermi liquid
triplet parameter $F>-1$. The precise value of $F$ is important for
the strength of spin thermalization caused by the electron-electron
interactions: as shown below, close to the Stoner instability at
$F=-1$ spin thermalization is quite strong, whereas it is much
weaker for $F\approx 0$.

The inter-spin energy exchange can be described by calculating the
spin thermalization heat current
\begin{equation}
\dot{Q}=\nu_{F}\Omega\int dE E I^{\uparrow \downarrow}(E),
\end{equation}
where $I^{\uparrow\downarrow}$ is the collision integral for
scattering between spin $\uparrow$ and spin $\downarrow$ electrons,
$\nu_F$ is the density of states at the Fermi level and $\Omega$ is
the volume of the spin valve spacer.

In the following we first discuss the resulting inter-spin
thermalization due to the regular spin conserving term. This can be
done for an arbitrary spin heat accumulation
$T_\uparrow-T_\downarrow$, where $T_{\uparrow(\downarrow)}$ is the
effective electron temperature for spin $\uparrow (\downarrow)$
electrons. Then we turn to the spin exchange term, concentrating on
the linear response regime in which $T_\uparrow-T_\downarrow \ll
(T_\uparrow+T_\downarrow)/2$. Within the calculation, we limit
ourselves to the case in which the difference between the chemical
potentials of the two spin species, {\it i.e.}, the spin
accumulation in the spacer $\mu_s=\mu_\uparrow-\mu_\downarrow
\approx 0$. This is because in the limit $\mu_s \ll T$ the effect of
a finite $\mu_s$ on the inter-spin relaxation is quadratic in
$\mu_s$.

For the electron-electron interaction, the effective dimensionality
(0D, 1D, 2D or 3D) of the spin valve island depends on the ratio
between the island thickness $d$, width $W$ and length $L$ to the
thermal coherence length $\xi_T$ (without losing generality, we
assume $d < W < L$). Alternatively, we can compare the temperature
to the Thouless energies $\hbar D/(L,W,d)^2$ defined by these size
scales. Therefore, 0d case is realized when $k_B T \ll \hbar D/L^2$,
the 1D case when $\hbar D/L^2 \ll k_B T \ll \hbar D/W^2$, the 2D
case when $\hbar D/W^2 \ll k_B T \ll \hbar D/d^2$ and the 3D case
when $k_B T \gg \hbar D/d^2$.

\section{Spin conserving term}
We assume that the spin-dependent electron distribution functions
can be written in terms of the Fermi-Dirac functions with
spin-dependent chemical potential $\mu_\sigma$ and temperature
$T_\sigma$ \cite{heikkilaup09}, i.e.,
$f_\sigma(E)=f^0(E;\mu_\sigma,T_\sigma)=\{1+\exp[(E-\mu_\sigma)/(k_B
T_\sigma)]\}^{-1}$. In this case, we can use the standard
relations\footnote{We set $\hbar=k_B=1$ in the intermediate results,
introducing them only in the final results.}
\begin{subequations}
\begin{align}
\begin{split}
&f^0(E;\mu,T) (1-f^0(E+x;\mu,T)) \\&=
[f^0(E+x;\mu,T)-f^0(E;\mu,T)]n(-x;T),
\end{split}\\
&\int dE [f^0(E;\mu,T)-f^0(E+x;\mu,T)]=x,
\end{align}
\end{subequations}
where $n(x,T)=\{1-\exp[x/(k_B T)]\}^{-1}$ is the Bose distribution
function. In the following, we use the short-hand notation
$n_\sigma(x)=n(x;T_\sigma)$. Using these relations, the spin
conserving collision integral can be written in the form
\begin{equation}
\begin{split}
I_b =& \frac{1}{2}\int d\omega \omega {\cal K}_b(\omega)
[f_\uparrow(E)-f_\uparrow(E-\omega)]\\
&\times
\underbrace{\left[\coth\left(\frac{\omega}{2k_BT_\downarrow}\right)-\coth\left(\frac{\omega}{2k_BT_\uparrow}\right)\right]}_{2[n_\uparrow(\omega)n_\downarrow(-\omega)-n_\uparrow(-\omega)n_\downarrow(\omega)]}.
\end{split}
\label{eq:spinconservingI}
\end{equation}
Here we used the symmetry of the kernel ${\cal K}_b(\omega)={\cal
K}_b(-\omega)$, by which the terms proportional to the chemical
potentials $\mu_\sigma$ vanish. With a kernel ${\cal K}_b(\omega) =
\kappa_b |\omega|^\alpha$, this yields for the spin conserving
inter-spin heat current
\begin{equation}
\dot{Q}_b =\frac{\nu_F \Omega \kappa_b \Gamma(4+\alpha) {\rm
Li}_{4+\alpha}(1) k_B^{4+\alpha}}{\hbar^{\alpha+3}}
(T_\uparrow^{4+\alpha}-T_\downarrow^{4+\alpha}),
\label{eq:spinconservingQ}
\end{equation}
where $\Gamma(x)$ is the gamma function and ${\rm Li}_n(x)$ is the
polylogarithm function. Using the ${\cal K}_b(\omega)$ from
Eqs.~\eqref{eq:1Dkernels}--\eqref{eq:3Dkernels}, we get
\begin{subequations}
\begin{align}
\dot{Q}_b^{\rm (1D)}&=\frac{3 \zeta \left(\frac{5}{2}\right)
k_B^{5/2}L }{8 \sqrt{2 \pi }  \hbar ^{3/2} \sqrt{D
(F+1)}}\left(T_\uparrow^{5/2}-T_\downarrow^{5/2}\right), \label{eq:Kb1D}\\
\dot{Q}_b^{\rm (2D)}&=\frac{\zeta (3) k_B^3 A}{4 \pi  D
(F+1)  \hbar ^2}\left(T_\uparrow^3-T_\downarrow^3\right), \label{eq:Kb2D}\\
\dot{Q}_b^{\rm (3D)}&=\frac{15 \zeta \left(\frac{7}{2}\right)
k_B^{7/2} \Omega}{32 \sqrt{2}\pi ^{3/2} \hbar ^{5/2} (D
(F+1))^{3/2}}\left(T_\uparrow^{7/2}-T_\downarrow^{7/2}\right),
\end{align}
\end{subequations}
where $\zeta(x)$ is the Riemann zeta-function and the superscript of
$\dot{Q}$ indicates the dimensionality. This term is regular, but
can in most cases be neglected since in 3D wires the spin
thermalization due to electron-electron interactions is typically
weaker than either the direct spin-flip scattering (at low
temperatures) or electron-phonon scattering (at high temperatures)
\cite{heikkilaup09}.

\section{Spin exchange term}
For the spin exchange term, the collision integral can be simplified
to
\begin{equation}
\begin{split}
&I_c = \frac{1}{2}\int d\omega d E' {\cal
K}_c(\omega)(f_\uparrow(E-\omega)-f_\uparrow(E'-\omega))\\&\times
(f_\downarrow(E')-f_\downarrow(E))
\underbrace{\left[\coth\left(\frac{E'-E}{2k_BT_\downarrow}\right)-\coth\left(\frac{E'-E}{2k_BT_\uparrow}\right)\right]}_{2[n_\uparrow(E'-E)n_\downarrow(E-E')-n_\uparrow(E-E')n_\downarrow(E'-E)]},
\end{split}
\end{equation}
leading to the inter-spin heat current
\begin{equation}
\begin{split}
&\dot{Q}_c=\frac{\nu_F \Omega}{8}\times \\&\int \frac{dE E dE'
d\omega {\cal
K}_c(\omega)\sinh\left(\frac{(T_\downarrow-T_\uparrow)(E'-E)}{2
T_\downarrow T_\uparrow}\right)}{\cosh\left(\frac{E+\omega}{2
T_\downarrow}\right) \cosh\left(\frac{E'+\omega}{2
T_\downarrow}\right)\cosh\left(\frac{E-\mu_s}{2 T_\uparrow}\right)
\cosh\left(\frac{E'-\mu_s}{2 T_\uparrow}\right)}.
\end{split}
\end{equation}
This integral is more difficult to handle than
Eq.~\eqref{eq:spinconservingI}. Therefore, we concentrate on the
linear response limit $\dot{Q}_c = K_c (T_\uparrow-T_\downarrow)$
and get
\begin{equation}
K_c=\frac{\nu_F \Omega}{16} T^3 \int \frac{dx dy dw {\cal
K}_c(Tw)x(x-y)}{\cosh\left(\frac{x}{2}\right)
\cosh\left(\frac{y}{2}\right)\cosh\left(\frac{x+w}{2}\right)\cosh\left(\frac{y+w}{2}\right)}.
\label{eq:Gc1}
\end{equation}
The integral over $x$ and $y$ can be carried out analytically,
leaving
\begin{equation}
K_c = \frac{\nu_F \Omega T^3}{24} \int_0^{\infty} dw \frac{w^2 {\cal
K}_c(Tw)(4\pi^2+w^2)}{\sinh^2\left(\frac{w}{2}\right)}.
\label{eq:Gc}
\end{equation}
In the 3D case with ${\cal K}_c \propto |\omega|^{-1/2}$, this
integral is regular and yields the result presented in
Ref.~\cite{heikkilaup09}. However, in the 1D and 2D cases the
integral over the first term has an infrared divergence and has to
be regularized. In the following we present a simple regularization
scheme based on the finite size of the sample.

Note that in Eq.~\eqref{eq:Gc} the cutoff scheme needs to be invoked
only for the first term. We include also the second term in the 2D
case, where the divergence is only logarithmic.

\subsection{1D wire}
As discussed in the Appendix, in finite systems the kernels actually
are sums over momenta. In the 1D wire the sum over the quantized
momenta $q_n=2\pi n/L$, where $L$ is the length of the wire, can be
carried out directly. In this case the characteristic scale for the
frequency is given by the Thouless energy $E_T^{\rm 1D}=D/L^2$.
When $|\omega| \gg E_T^{\rm 1D}$, we recover
Eq.~\eqref{eq:1Dkernels}. For $|\omega| \ll E_T^{\rm 1D}$, the
kernel does not depend on $\omega$,
\begin{equation}
{\cal K}_c (|\omega| \ll E_T^{\rm 1D}) = \frac{\pi^2 F^2}{180 \Omega
(E_T^{\rm 1D})^2 (1+F)^2}.
\end{equation}
We can hence use $E_T^{\rm 1D}$ as a lower cutoff in
Eq.~\eqref{eq:Gc}. In principle, we should also include the part of
the kernel which is constant for low frequencies $\omega \lesssim
E_T^{\rm 1D}$. However, the contribution to the integral from this
part is of the order of $(E_T^{\rm 1D}/T)^{3/2}$ times the
contribution of the second part, and can hence be disregarded in the
1D limit $E_T^{\rm 1D} \ll T$. The remaining integral is $K_c\propto
I_1$ with
\begin{equation}
I_1 = \int_{E_T^{\rm 1D}/T}^\infty \frac{\sqrt{w} dw}{\sinh^2(w/2)}.
\end{equation}
In the limit $k_B T \gg E_T^{\rm 1D}$ the most significant
contribution comes from frequencies $w \ll 1$ for which $\sinh(w)
\approx w$, leading to
\begin{equation}
I_1 \approx 8 \sqrt{\frac{T}{E_T^{\rm 1D}}}.
\end{equation}
Including this contribution, we obtain for the spin thermalization
heat conductance from the spin exchange contribution
\begin{equation}
K_c^{\rm 1D} = \frac{F^2 k_B 4\sqrt{2} \pi
   (k_B T)^2}{3 \hbar E_T^{\rm 1D} (F+2)
   \left(F+\sqrt{F+1}+1\right)},
   \label{eq:Kc1D}
   \end{equation}
where we reintroduced $k_B$ and $\hbar$. The diverging term hence
gives rise to a temperature dependent scaling as $\propto T^2$.

As discussed in Ref.~\cite{heikkilaup09}, a spin valve can be
characterized in terms of the temperature above which spin
thermalization is stronger than heat diffusion through the contacts,
i.e., $K_c \gg \Lor G_0 T$, where $G_0$ is the spin-averaged contact
conductance and $\Lor=\pi^2 k_B^2/(3e^2)$ is the Lorenz number. This
characteristic temperature for electron-electron interaction in a 1D
sample is
\begin{equation}
T_{\rm ch,e-e}^{\rm 1D}=\frac{ (F+2) \left(F+\sqrt{F+1}+1\right)}{16
\sqrt{2} F^2} \frac{E_T^{\rm 1D}}{k_B} g,
\end{equation}
where $g=G_0/(e^2/h)$ is the dimensionless conductance of the
contacts. Let us estimate the characteristic temperature and the
thermal coherence length $\xi_T$. The latter is
\begin{equation}
\xi_T \approx 85 \text{ nm} \times \left(\frac{D}{0.001 \text{
m$^2$/s}} \frac{1 \text{ K}}{T}\right)^{1/2}.
\end{equation}
Wires with lateral dimensions less than $\xi_T$ fall into the 1D
limit. Connecting such wires to reservoirs via contacts with
resistance 10 $\Omega$ and assuming $F=-0.3$, we get
\begin{equation}
T_{\rm ch,e-e}^{\rm 1D} \overset{F=-0.3}{\approx} 25 \text{ K}
\times \left(\frac{D}{0.001 \text{ m$^2$/s}} \left(\frac{1 \text{
}\mu\text{m}}{L}\right)^2 \frac{G_0}{0.1 \text{ S}} \right).
\end{equation}
For wires fabricated from materials close to the Stoner instability,
such as Pd, say with $F=-0.9$, this characteristic temperature is
\begin{equation*}
T_{\rm ch,e-e}^{\rm 1D} \overset{F=-0.9}{\approx} 0.5 \text{ K}
\times \left(\frac{D}{0.001 \text{ m$^2$/s}} \left(\frac{1 \text{
}\mu\text{m}}{L}\right)^2 \frac{G_0}{0.1 \text{ S}} \right).
\end{equation*}
These values should be compared to the characteristic temperature
due to the electron-phonon interaction using the results from
Ref.~\cite{heikkilaup09}. This is
\begin{equation}
T_{\rm ch,e-ph} \approx 4 \text{ K} \times \left(\frac{10^9 \text{
Wm$^{-3}$K$^{-5}$}}{\Sigma} \frac{0.015 \text{
($\mu$m)$^3$}}{\Omega} \frac{G_0}{0.1 \text{ S}}\right)^{-1/3}.
\end{equation}
In this case electron-electron interaction in systems close to the
Stoner instability is the dominating spin thermalization mechanism.

Another way to characterize the spin energy exchange is via the spin
thermalization time $\tau_{\rm st}=\Lor e^2 \nu_F T \Omega/(2 K_{\rm
e-e})$. From Eq.~\eqref{eq:Kc1D}:
\begin{equation*}
\tau_{\rm st,e-e}^{\rm 1D}=\frac{\pi   (F+2)
\left(F+\sqrt{F+1}+1\right) \hbar \nu_F \Omega }{8
   \sqrt{2} F^2 } \frac{E_T^{\rm 1D}}{k_B T}.
\end{equation*}
This scattering time should be compared to the direct spin-flip time
of roughly 100 ps in typical samples \cite{jedema02}. With $F=-0.3$
and some other typical values for metals with wire cross section
$A=Wd$, $\tau_{\rm st,e-e}$ is rather long,
\begin{equation*}
\begin{split}
\tau_{\rm st,e-e}^{\rm 1D}& \overset{F=-0.3}{\approx} 1 \text{ ns}
\\&\times \left[\frac{\nu_F}{10^{47} \text{ J$^{-1}$m$^{-3}$}}
\frac{A}{1500 \text{ nm$^2$}} \frac{1 \text{ }\mu\text{m}}{L}
\frac{D}{0.001 \text{ m$^2$/s}} \frac{1 \text{K}}{T}\right]
\end{split}
\end{equation*}
whereas for $F=-0.9$ it is
\begin{equation}
\begin{split}
\tau_{\rm st,e-e}^{\rm 1D} &\overset{F=-0.9}{\approx} 20 \text{ ps}
\\&\times \left[\frac{\nu_F}{10^{47} \text{ J$^{-1}$m$^{-3}$}}
\frac{A}{1500 \text{ nm$^2$}} \frac{1 \text{ }\mu\text{m}}{L}
\frac{D}{0.001 \text{ m$^2$/s}} \frac{1 \text{K}}{T}\right].
\end{split}
\end{equation}
These examples show that the electron-electron interaction in 1D
samples is especially relevant for systems close to the Stoner
instability.

\subsection{2D square sheet}
In the two-dimensional case ${\cal K}_c = \kappa_{2D} |\omega|^{-1}$
and the resulting integral has a logarithmic divergence. Now the
regular term in Eq.~\eqref{eq:Gc} yields
\begin{equation*}
\nu_F \Omega \zeta(3) \kappa_{2D} T^2
\end{equation*}
and for the diverging term, we need to evaluate the integral
\begin{equation}
\begin{split}
I_2 &= \int_{E_T/T}^\infty \frac{w dw}{\sinh^2(w/2)} \\&= 2 \frac{E_T}{T} \coth\left(\frac{E_T}{2 T}\right)-2 \ln\left[2\cosh\left(\frac{E_T}{T}\right)-1\right]
\\&\approx 4\left[\ln\left(\frac{T}{E_T}\right)+1\right].
\end{split}
\end{equation}
Here the lower cutoff $E_T=\hbar D/W^2$ is the confinement energy
due to the smaller of the two extended dimensions. Hence, for the
full spin thermalization heat conductance we get
\begin{equation}
K_c \overset{\rm 2D}{=} \frac{F^2 k_B^3 T^2 \left(4 \pi ^2 \left(\ln
\left(\frac{T}{E_T}\right)+1\right)+6 \zeta
   (3)\right)}{12 \pi \hbar E_T^{\rm 2D} \left(F^2+3 F+2\right)},
   \label{eq:Kc2D}
\end{equation}
where $E_T^{\rm 2D}=\hbar D/(LW)$.

In this case the characteristic temperature is
\begin{equation}
T_{\rm ch, e-e}^{\rm 2D}=\frac{(F+1) (F+2) g}{4 F^2
W\left(\frac{(F+1) (F+2) g e^{\frac{3 (F (2
   F+3)+6) \zeta (3)}{4 \pi ^2 F^2}+1}}{2 F^2} \frac{E_T^{\rm 2D}}{E_T} \right)} \frac{E_T^{\rm 2D}}{k_B},
\end{equation}
where $W(x)$ is the principal solution for $z$ in $x=ze^z$. Here we
included also the regular terms from Eqs.~\eqref{eq:Kb2D} and
\eqref{eq:Kc2D} besides the term proportional to $\ln(T/E_T)$. To
estimate this characteristic temperature, we consider a square bar
of area (width $\times$ length) $A=1$ ($\mu$m)$^2$ and much thinner
than $\xi_T$. In this case we have $T_{\rm ch,e-e}^{\rm 2D} \approx
10$ K for wires with $F=-0.3$ and $T_{\rm ch,e-e}^{\rm 2D} \approx
0.3$ K with $F=-0.9$ and otherwise similar values as in the above 1D
case. For comparison, with the thickness $d=30$ nm, we get for the
phonon contribution $T_{\rm ch, e-ph} \approx 3.2$ K.

The spin thermalization time is
\begin{equation}
\begin{split}
&\tau_{\rm st, e-e}^{\rm 2D}=\\&\frac{4 \pi ^3  (F+1) (F+2) \nu_F
\Omega \hbar }{ \left(8 \pi ^2
   F^2 \left(\log \left(\frac{k_B T}{E_T}\right)+1\right)+3 (2+F+4F^2) \zeta
   (3)\right)}\frac{E_T^{\rm 2D}}{k_B T}.
   \end{split}
   \end{equation}
Assuming a film thickness of $30$ nm, for $F=-0.3$ we get $\tau_{\rm
st, e-e} \approx 7$ ns whereas for $F=-0.9$ $\tau_{\rm st, e-e}
\approx 80$ ps at $T=1$ K and otherwise similar conditions as above.

\begin{figure}[h]
\centering
\includegraphics[width=\columnwidth]{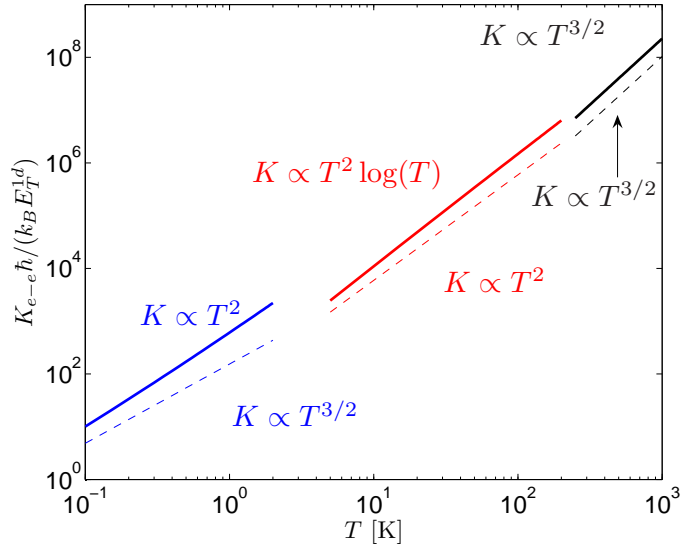}
\caption{(Color online): Temperature dependence of the spin
thermalization heat conductance due to electron-electron interaction
for a wire with length $1$ $\mu$m, width 100 nm and thickness 10 nm,
diffusion constant $D=0.03$ m$^2$/s and Fermi liquid parameter
$F=-0.3$ (solid lines) and $F=0$ (dashed lines). The three regimes
plotted in the figure are for the 1D, 2D and 3D limits. The heat
conductance is written in units of $k_B E_T^{\rm 1D}/\hbar$, where
$E_T^{\rm 1D}/k_B=22$ mK, corresponding to the crossover temperature
between 0d and 1D behavior. In the case $F=0$ the spin exchange
contribution vanishes, which changes the temperature dependences of
the heat conductances in the 1D and 2D cases. } \label{fig:KvsT}
\end{figure}

\section{Conclusions}

One of the key features enabling the success of spintronics in the
past two decades is the long spin-flip time found in metals. One
recent trend has been the study of thermal transport in spintronic
systems. Special interest has been devoted to the study of
magnetothermal effects, where the heat transport depends on the
magnetic configuration of the system. Besides the long spin
relaxation time, such effects rely on relatively weak inter-spin
energy relaxation. This was studied in detail by the present authors
in Ref.~\cite{heikkilaup09}. It was found, among other things, that
making the sample smaller reduces the effects of inter-spin
relaxation. However, the conventional 3D theory of electron-electron
interaction in disordered metals applies only when the system
dimensions exceed the thermal coherence length $\xi_T$. Moreover, it
was pointed out in Refs.~\cite{dimitrova07,chtchelkatchev08} that
the electron-electron contribution to the energy relaxation in
reduced dimensions diverges. In this paper we have addressed this
divergence by introducing a natural cutoff emerging from the sample
size. The main results of this consideration, Eqs.~\eqref{eq:Kc1D}
and \eqref{eq:Kc2D}, show that indeed the electron-electron
interactions in 1D and 2D samples are more relevant for the spin
thermalization than in the 3D case. However, for typical normal
metal spacers for which the Fermi liquid triplet parameter $F$ is
far from the Stoner instability $F=-1$, this effect is still masked
by direct spin-flip processes at low temperatures and
electron-phonon relaxation at high temperatures. The situation is
different for materials close to the Stoner instability, for example
palladium wires doped with nickel. In these systems the inter-spin
energy relaxation due to electron-electron interaction may be the
dominating relaxation mechanism.

The behavior of the spin thermalization heat conductance due to
electron-electron scattering is illustrated in Fig.~\ref{fig:KvsT}.
It shows the regimes of different dimensionality for an example
system.

The theory presented in this manuscript is based on the collision
integrals calculated in Ref.~\cite{dimitrova07}, which generalize
the Altshuler-Aronov theory \cite{altshuleraronov} to the
spin-dependent case. We point out that the electron-electron energy
relaxation described by this theory in the 1D limit has been
investigated experimentally in detail \cite{huard04}. However,
according to these experiments the measured energy relaxation is
stronger by roughly an order of magnitude than that predicted by the
theory. This is evidence that the results for the spin
thermalization conductances presented in Eqs.~\eqref{eq:Kb1D},
\eqref{eq:Kb2D}, \eqref{eq:Kc1D} and \eqref{eq:Kc2D} might be
underestimated.

\section*{Acknowledgments}

This work was supported by the Academy of Finland, the Finnish
Cultural Foundation, and NanoNed, a nanotechnology programme of the
Dutch Ministry of Economic Affairs. TTH acknowledges the hospitality
of Delft University of Technology, where this work was initiated.

\appendix

\section{Collision integrals and kernels}
The collision integrals describing electron-electron scattering in
spin-dependent systems are
\begin{subequations}
\begin{align}
\begin{split}
&I_a(E) = \int d\omega dE' {\cal K}_a(\omega)\times\\
&[(1-f_\sigma(E))(1-f_{\sigma}(E'))f_\sigma(E-\omega)f_{\sigma}(E'+\omega)\\&-f_\sigma(E)
f_{\sigma}(E') (1-f_\sigma(E- \omega))(1-f_{\sigma}(E'+ \omega))]
\end{split}\\
\begin{split}
&I_b(E) = \int d\omega dE' {\cal K}_b(\omega)\times\\
&[(1-f_\sigma(E))(1-f_{-\sigma}(E'))f_\sigma(E-\omega)f_{-\sigma}(E'+\omega)\\&-f_\sigma(E)
f_{-\sigma}(E') (1-f_\sigma(E- \omega))(1-f_{-\sigma}(E'+ \omega))]
\end{split}\\
\begin{split}
&I_c(E) = \int d\omega dE' {\cal K}_c(\omega)\times\\
&[(1-f_\sigma(E))(1-f_{-\sigma}(E'))f_\sigma(E'-\omega)f_{-\sigma}(E+\omega)\\&-f_\sigma(E)
f_{-\sigma}(E') (1-f_\sigma(E'- \omega))(1-f_{-\sigma}(E+ \omega))].
\end{split}
\end{align}
\end{subequations}
The kernels $K_{a,b,c}(\omega)$ depend on the dimensionality of the
sample \cite{dimitrova07}:
\begin{subequations}
\begin{align}
{\cal K}_{a}&=\frac{8}{2\pi \nu_F \Omega} \sum_{\q}
\frac{1}{\omega^2+(D
\q^2)^2}\frac{\left(\frac{1}{2}+F\right)^2+\frac{\omega}{(2 D
\q^2)^2}}{(1+F)^2+\frac{\omega^2}{(D\q^2)^2}}\\
{\cal K}_{b}&=\frac{8}{2\pi \nu_F \Omega} \sum_{\q}
\frac{1}{\omega^2+(D \q^2)^2}\frac{\frac{1}{4}+\frac{\omega}{(2 D
\q^2)^2}}{(1+F)^2+\frac{\omega^2}{(D\q^2)^2}}\\
{\cal K}_{c}&=\frac{8}{2\pi \nu_F \Omega} \sum_{\q}
\frac{1}{\omega^2+(D \q^2)^2}\frac{F^2}{(1+F)^2+\frac{(\omega-F
\mu_s)^2}{(D\q^2)^2}}.
\end{align}
\end{subequations}
Here $D$ is the diffusion constant in the spacer and $F$ the Fermi
liquid interaction parameter in the triplet channel.

In this paper we describe systems with at least one extended
dimension, {\it i.e.}, quasi-1D, quasi-2D or 3D metal islands.
According to the standard prescriptions for calculating the sums
\begin{align*}
\sum_\q \mapsto L/(\pi) \int_0^\infty dq,& \quad d=1\\
\sum_\q \mapsto A/(2\pi) \int_0^\infty q dq,& \quad d=2\\
\sum_\q \mapsto \Omega/(2\pi^2) \int_0^\infty q^2 dq, & \quad d=3,
\end{align*}
we get for $d=1$
\begin{subequations}
\begin{align}
{\cal K}_a&=\frac{\frac{4 F (F+1)}{(F+2)
\left(\sqrt{F+1}+1\right)}+1}{2
 \pi  A \nu_F  \sqrt{2D (F+1)}} \frac{1}{|\omega|^{3/2}}\\
{\cal K}_b&=\frac{1}{2  \pi  A \nu_F  \sqrt{2D
(F+1)}}\frac{1}{|\omega|^{3/2}}\\
\begin{split}
{\cal K}_c &=\frac{\sqrt{2} F^2}{\pi  A \nu_F  \sqrt{D (F+1)} {\cal
C}}\\&\overset{\mu_s \rightarrow 0}{\rightarrow} \frac{\sqrt{2}
F^2}{\pi  A (F+2) \left(\sqrt{F+1}+1\right) \nu_F |\omega |^{3/2}
\sqrt{D (F+1)}}
\end{split}
\end{align}
\label{eq:1Dkernels}
\end{subequations}
Here ${\cal C}=(\sqrt{|\omega-F \mu_s |}+\sqrt{F+1} \sqrt{|\omega|})
(|\omega-F \mu_s |+(F+1) |\omega|)$. This result was also obtained
in Ref.~\cite{dimitrova07}. For $d=2$ the kernels are
\begin{subequations}
\begin{align}
{\cal K}_a &= \frac{1+2 F^2+\frac{5 F}{2}}{\pi  D \left(4 F^2+12
F+8\right)
d \nu_F  |\omega|}\\
{\cal K}_b &= \frac{1}{8 \pi  D (F+1) d \nu_F  |\omega |}\\
\begin{split}
{\cal K}_c &= \frac{F^2}{2 \pi  D (F+1) d \nu_F  (|\omega -F \mu_s
|+(F+1) |\omega |)} \\&\overset{\mu_s \rightarrow 0}{\rightarrow}
\frac{F^2}{2 \pi D \left(F^2+3 F+2\right) d \nu_F  |\omega |}
\end{split}
\end{align}
\label{eq:2Dkernels}
\end{subequations}
and for $d=3$ we get
\begin{subequations}
\begin{align}
{\cal K}_a &= \frac{1+\frac{4 (F+1) \left(\sqrt{F+1} F+\sqrt{F+1}-1\right)}{F+2}}{2 \pi ^2 \nu_F  \sqrt{|\omega |} (2D (F+1))^{3/2}}\\
{\cal K}_b &= \frac{1}{2 \pi ^2 \nu_F  \sqrt{|\omega |} (2D
(F+1))^{3/2}}\\
\begin{split}
{\cal K}_c &= \frac{F \left(|(F+1) \omega |^{3/2}-|\omega -F  \mu_s
|^{3/2}\right)}{\sqrt{2} \pi ^2 \nu_F  (D (F+1))^{3/2} (\mu_s
+\omega ) ((F+2)
   \omega -F \mu_s )}\\& \overset{\mu_s \rightarrow 0}{\rightarrow} \frac{F
\left((F+1)^2-\sqrt{F+1}\right)}{\sqrt{2} \pi ^2 D^{3/2} (F+1)^2
(F+2) \nu_F  \sqrt{|\omega |}}.
\end{split}
\end{align}
\label{eq:3Dkernels}
\end{subequations}
In the present paper, we only use the kernels with $\mu_s\rightarrow
0$. This is because in the limit $\mu_s \ll T$ the effect of a
finite $\mu_s$ is quadratic in $\mu_s$.


\section*{References}


\begin{thebibliography}{1}
\expandafter\ifx\csname url\endcsname\relax
  \def\url#1{\texttt{#1}}\fi
\expandafter\ifx\csname urlprefix\endcsname\relax\def\urlprefix{URL
}\fi \expandafter\ifx\csname href\endcsname\relax
  \def\href#1#2{#2} \def\path#1{#1}\fi

\bibitem{hatami07}
M.~Hatami, G.~E.~W. Bauer, Q.~Zhang, P.~J. Kelly, Phys. Rev. Lett.
99(6) (2007) 066603.

\bibitem{hatami09}
M.~Hatami, G.~E.~W. Bauer, Q.~Zhang, P.~J. Kelly, Phys. Rev. B 79
(2009) 174426.

\bibitem{dubi09}
Y.~Dubi, M.~D. Ventra, Phys. Rev. B 79 (2009) 081302.

\bibitem{heikkilaup09}
T.~T. Heikkil\"a, M.~Hatami, G.~E.~W. Bauer, Phys. Rev. B 81, (2010) 100408(R).

\bibitem{dimitrova07}
O.~Dimitrova, V.~Kravtsov,  JETP Lett. 86 (2007) 670.

\bibitem{chtchelkatchev08}
N.~M. Chtchelkatchev, I.~S. Burmistrov, Phys. Rev. Lett. 100 (2008)
206804.

\bibitem{jedema02}
F.~J. Jedema, H.~B. Heersche, A.~T. Filip, J.~J.~A. Baselmans, B.~J.
van Wees,
  Nature 416 (2002) 713.

\bibitem{altshuleraronov}
B.~Altshuler, A.~Aronov, in: A.~Efros, M.~Pollak (Eds.),
Electron-Electron
  Interactions in Disordered Systems, Elsevier, Amsterdam, 1985.

\bibitem{huard04}
B.~Huard, A.~Anthore, F.~Pierre, H.~Pothier, N.~O. Birge,
D.~Esteve",
  Solid State Commun. 131 (2004) 599 -- 607.

\end{thebibliography}
\end{document}